%% file: main.tex
\documentclass[10pt,twocolumn,letterpaper]{article}

\usepackage{iccv}
\usepackage{times}
\usepackage{epsfig}
\usepackage{graphicx}
\usepackage{amsmath}
\usepackage{amssymb}
\usepackage{caption}

\usepackage{xcolor}

\usepackage[pagebackref=true,breaklinks=true,letterpaper=true,colorlinks,bookmarks=false]{hyperref}

\iccvfinalcopy 

\def\iccvPaperID{7327} 

\ificcvfinal\fi

\begin{document}

\title{The Power of Sound (TPoS):\\
Audio Reactive Video Generation with Stable Diffusion}


\author{Your Name Here}

\author{Yujin Jeong\textsuperscript{1}, Wonjeong Ryoo\textsuperscript{2}, Seunghyun Lee\textsuperscript{2}, Dabin Seo\textsuperscript{1}, \\ Wonmin Byeon\textsuperscript{3},
Sangpil Kim,\textsuperscript{2,*} and Jinkyu Kim\textsuperscript{1,*} \\
\textsuperscript{1}~Department of Computer Science and Engineering, Korea University, Seoul 02841, Korea \\
\textsuperscript{2}~Department of Artificial Intelligence, Korea University, Seoul 02841, Korea \\
\textsuperscript{3}~NVIDIA Research, Santa Clara 95050, USA\\
{\small \textsuperscript{*}Correspondences: S. Kim ({\tt spk7@korea.ac.kr}) and J. Kim ({\tt jinkyukim@korea.ac.kr})}
}

\twocolumn[{%
 \renewcommand\twocolumn[1][]{#1}%
\maketitle
\vspace{-12mm}
  \begin{center}
\centering
    \captionsetup{type=figure}
\includegraphics[width=.88\linewidth]{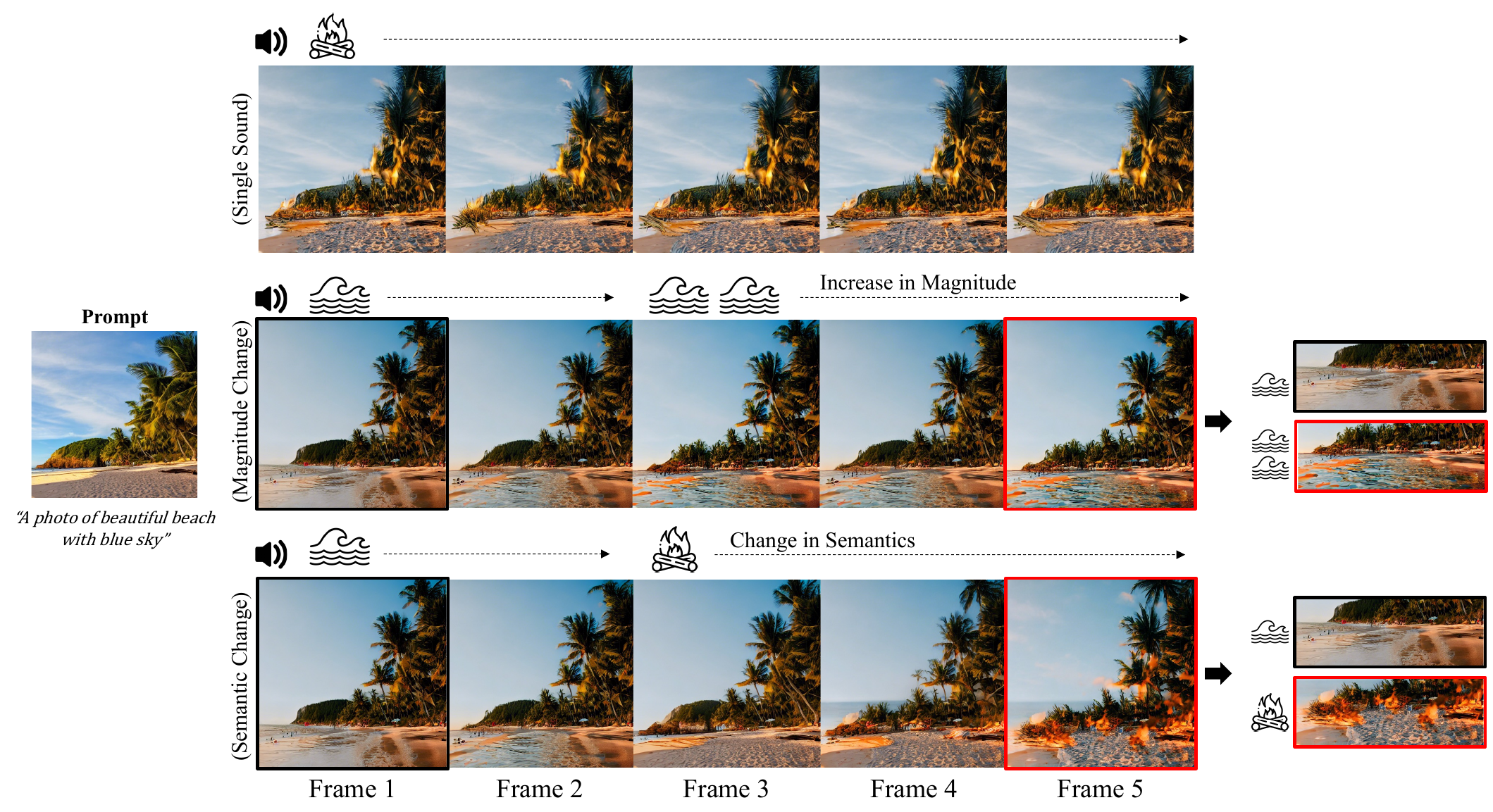} 
  \captionof{figure}{The Power of Sound (TPoS) is a novel framework that generates audio-reactive video sequences. Built upon the Stable Diffusion model, our model first generates an initial frame from a user-provided text prompt (e.g. ``a photo of a beautiful beach with a blue sky''), then reactively manipulates the style of generated images corresponding to the sound inputs (e.g. an audio sequence of fireplace). Our model is indeed able to generate a frame conditioned on semantic information of the sound (see 1st and 2nd rows where images are manipulated driven by sound inputs such as fireplace or wave sound), while realistically dealing with temporal visual changes conditioned on changes of sound, e.g., increasing magnitude of sounds (see second row) or wave $\rightarrow$ fireplace (see last row). TPoS creates visually compelling and contextually relevant video sequences in an open domain. 
}
\end{center}%
 }]

\ificcvfinal\thispagestyle{empty}\fi

\input{src/abstract.tex}


\input{src/introduction.tex}
\input{src/related_work.tex}

\input{src/method.tex}

\input{src/experiments.tex}
\input{src/application.tex}
\input{src/conclusion.tex}

\myparagraph{Acknowledgements.}
{\small This work was supported by the National Research Foundation of Korea grant (NRF-2021R1C1C1009608, 10\%), Basic Science Research Program (NRF-2021R1A6A1A13044830, 5\%), Institute of Information \& communications Technology Planning \& Evaluation (IITP) grant funded by the Korea government(2022-0-00043, 30\%), and the MSIT(Ministry of Science and ICT), Korea, under the ICT Creative Consilience program(IITP-2023-2020-0-01819, 5\%) supervised by the IITP. W. Ryoo, S. Lee, and S. Kim are supported by the Institute of Information \& communications Technology Planning \& Evaluation (IITP) grant funded by the Korea government(MSIT) (No. 2019-0-00079, Artificial Intelligence Graduate School Program(Korea University), 30\%), the National Research Foundation of Korea grant (NRF-2022R1F1A1074334, 15\%),  Culture, Sports, and Tourism R\&D Program through the Korea Creative Content Agency grant funded by the Ministry of Culture, Sports and Tourism in 2023 (Project Name: 4D Content Generation and Copyright Protection with Artificial Intelligence, Project Number: R2022020068, 5\%), and supported by the Google Cloud Research Credits program(code- MU6X-FKAV-N3A3-1X0T).}

\clearpage
{\small
\bibliographystyle{ieee_fullname}
\bibliography{egbib}
}




\end{document}

%% file: src/abstract.tex
\begin{abstract}
In recent years, video generation has become a prominent generative tool and has drawn significant attention. However, there is little consideration in audio-to-video generation, though audio contains unique qualities like temporal semantics and magnitude. Hence, we propose The Power of Sound (TPoS) model to incorporate audio input that includes both changeable temporal semantics and magnitude. To generate video frames, TPoS utilizes a latent stable diffusion model with textual semantic information, which is then guided by the sequential audio embedding from our pretrained Audio Encoder. As a result, this method produces audio reactive video contents. We demonstrate the effectiveness of TPoS across various tasks and compare its results with current state-of-the-art techniques in the field of audio-to-video generation. 
More examples are available at \url{https://ku-vai.github.io/TPoS/}
\end{abstract}

%% file: src/introduction.tex
\section{Introduction}
Recent generative models have demonstrated the potential to generate visually-appealing video frames~\cite{singer2022make, ho2022video, ho2022imagen, molad2023dreamix, villegas2022phenaki}. They often use a simple text prompt (e.g., ``a video of a person on the street on a rainy day'') to generate a video which is intuitive for end-users to drive a video generation. Text can effectively convey uni-modal object-wise guidance, such as ``a rainy day'' or ``a person'', but it may be challenging if users want to drive it into more complex sequential procedures, i.e., ``a video of a person on the street on a rainy day, but a rain suddenly stops, and a wind blows.''




In this paper, we leverage sounds to guide the video generation models, i.e., sound-driven video generation. Audio is another modality that can complement texts by effectively providing sequential information (or temporal semantics): e.g., a continuous transition from the sound of light rain to the sound of heavy rain. There have been introduced sound-guided video generation approaches. However, existing sound-guided video generation approaches are limited to specific applications, such as face generation~\cite{ji2021audio, prajwal2020lip, liang2022expressive, ji2022eamm, liang2022expressive}, where audio is used to provide a script for the avatar, or other simple synthetic motions (e.g., a video of musicians playing violin or a video of painting motions by an artist)~\cite{gross2019automatic, jeong2021traumerai, le2021ccvs, chatterjee2020sound2sight}. 

Recently, Lee~\etal~\cite{lee2022eccv} introduced a sound-guided landscape video generation model, leveraging the latent space of StyleGAN~\cite{karras2021alias}. They focus on using audio only for semantic labels (i.e., a sound of the wind is simply encoded into a meaning of wind) but not temporal semantics -- i.e. semantic information that changes over time. Thus, in this work, we focus on leveraging temporal semantics from audio inputs such that our video generator reactively manipulates video frames. Our model temporally aligns the latent space with the given audio sequence (e.g., continuous changes in audio, e.g., weak rain $\rightarrow$ heavy rain, are reflected to generate corresponding video frames). 


Our work starts with Stable Diffusion~\cite{rombach2022high}, a text-driven image generator with advantages in generating high-resolution images based on the latent diffusion models. Its architectural advantages (i.e., attention mechanism and diffusion process) help leverage audio as a driving condition, generating temporally reactive and consistent video frames. Given the latent space of trained Stable Diffusion, we generate video frames temporally guided by audio sequences with regularizers to ensure temporal consistency (between generated consecutive frames) and correspondence with audio inputs.

\begin{figure}
\begin{center}
      \includegraphics[width=\linewidth]{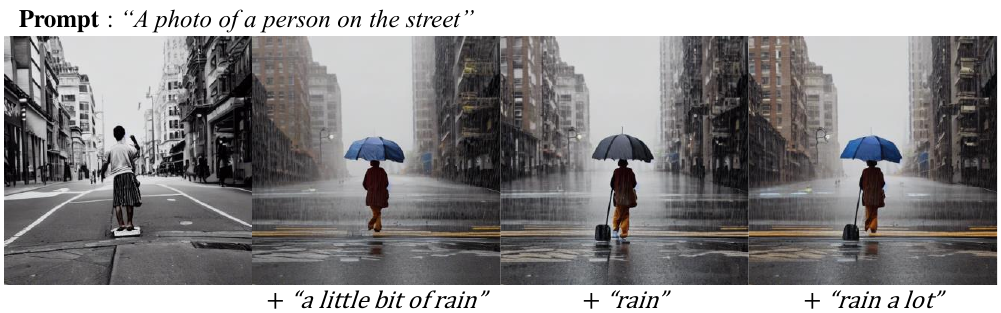} %
  \caption{Limitation of text-driven image manipulation. Given a generated image by Stable Diffusion~\cite{rombach2022high} with a text prompt, ``A photo of a person on the street,'' additional textual conditions of different semantic meanings (i.e. ``a little of rain'', ``rain'', and ``rain a lot'') produce similar images, failing to capture differences. Note that we apply SEGA~\cite{brack2022stable}'s guidance to preserve content identity.}
  \label{fig:text}
  \vspace{-1.5em}
\end{center}
\end{figure} 

Our model consists of two main modules: (i) {\em Audio Encoder}, which is designed to encode temporal semantics of audio sequences, producing a sequence of the latent vectors. (ii) {\em Audio Semantic Guidance Module}, which uses the above-mentioned latent vectors as a condition in the diffusion process to generate corresponding image outputs. We apply identity regularizer to produce temporally consistent video frames, while we apply audio semantic guidance to generate audio-reactive video frames. 

However, we observe that generating multifarious high-quality images solely from a sound input is challenging due to the lack of such a large-scale dataset to train a model. Instead, we first generate an initial frame using pre-trained Stable Diffusion model with a text prompt, then generate the following video frames conditioned on audio inputs. This frees a model from data dependency burdens and enables training with the current relatively small (than image-text modality) audio dataset~\cite{lee2022eccv, chen2020vggsound}
, focusing on leveraging temporal semantics from audio. We summarize our contributions as follows:
\vspace{-0.5em}
\begin{itemize} 
    \item We propose a novel sound-driven video generation method built upon Stable Diffusion~\cite{rombach2022high} and can generate video frames reactively with audio sequence inputs. \vspace{-1.75em} 
    \item Our attention-based Audio Encoder produces temporally-aware latent vectors, which are consumed by Stable Diffusion as a per-time manipulation condition, producing audio-reactive video frames. \vspace{-0.5em}
    \item Our model regularizes the latent features of diffusion models to produce temporally consistent video frames, preserving identity throughout the generated video.\vspace{-0.5em}
    \item We demonstrate the effectiveness of our proposed model using a public dataset Landscape~\cite{lee2022eccv}, generally outperforming other state-of-the-art sound-driven video generation approaches in terms of video quality metrics and human evaluation. 
\end{itemize}

%% file: src/related_work.tex
\section{Related Work}


\myparagraph{Latent Diffusion Models.}
Recent success~\cite{rombach2022high} suggests that the Latent Diffusion Models (LDM) improve the efficiency of the diffusion process, successfully generating high-quality images given a text prompt. One challenge in LDM is that the generation process is too sensitive to the condition, making it difficult to control semantics. Recently, there have been introduced to control semantics with LDM by a semantic mask~\cite{avrahami2022blended} or by utilizing semantic information of the cross-attention layers~\cite{hertz2022prompt}. Wu~\etal~\cite{wu2022uncovering} used linear combinations of text embeddings and Liu~\etal~\cite{liu2022compositional} proposed composable diffusion models, but they still remained challenging to control fine-grained semantic changes. Recently, Semantic Guidance (SEGA)~\cite{brack2022stable} computed a guidance vector in the latent space, enabling semantic control of diffusion models without further inputs. Inspired by SEGA, we also control the semantics in the latent space with temporally-encoded audio vector sequences.

\input{src/overview}

\myparagraph{Text-driven Video Generation.}
Recent text-to-video generation tools, including Make-A-Video~\cite{singer2022make}, Video Diffusion Models~\cite{ho2022video}, Imagen video~\cite{ho2022imagen}, and Phenaki~\cite{villegas2022phenaki} have shown promising performance in generating videos from textual descriptions. However, text-to-video generation has its limitations in terms of temporal coherence, which mostly leads to short video duration or a linear video change. Recent text-to-video generation methods, StyleGAN-V~\cite{skorokhodov2022stylegan} and Dreamix~\cite{molad2023dreamix}, made progress addressing these issues. However, conditioning temporal semantics or complex scenarios is still challenging to be obtained from text inputs. Thus, in this paper, we want to explore conditioning a model with audio inputs, which inherently convey such temporal semantics.

\begin{table}[t]
    \centering
    \setlength{\tabcolsep}{6pt}
    \renewcommand{\arraystretch}{1.3} 
    \caption{Comparison between existing state-of-the-art audio-driven video generation approaches in terms of whether they consider the following factors: temporal semantics, magnitude changes of sound, and target domains.} \vspace{-.5em}
    \label{tab:Comparison}
    \resizebox{\linewidth}{!}{
        \begin{tabular}{@{}lccl}\toprule
            Model & Temporal Semantics & Magnitude & Domains (Audio Type) \\\midrule
            Sound2Sight~\cite{chatterjee2020sound2sight} & - & \checkmark & Closed  \\ 
            CCVS~\cite{le2021ccvs} & - & \checkmark & Closed (Music) \\ 
            Tr\"aumerAI~\cite{jeong2021traumerai}  & - & \checkmark  & Closed (Music)  \\ 
            Lee~\etal~\cite{lee2022eccv} & \checkmark & - & Closed (Nature) \\\midrule
            Ours & \checkmark & \checkmark & Open Domains \\\bottomrule
        \end{tabular}
        }\vspace{-1.5em}
\end{table}

\myparagraph{Audio-driven Video Generation.}
Leveraging temporal semantics was not seriously considered in previous audio-driven video generation approaches. Sound2Sight\cite{chatterjee2020sound2sight} and CCVS~\cite{le2021ccvs} generate video frames conditioned on the (non-temporal) context of the given audio, while Tr\"aumerAI~\cite{jeong2021traumerai} utilized the magnitude of the given audio. Recently, Lee~\etal~\cite{lee2022eccv} explored a model that can consider audio semantics as a condition to drive a video generator. Also, their dependency on StyleGAN~\cite{skorokhodov2022stylegan}-based embedding space makes it difficult for models to generate transitions in video. In this work, we focus on leveraging temporal semantics from audio inputs such that our generator reactively manipulates video frames.

%% file: src/overview.tex
\begin{figure*}
\begin{center}
      \includegraphics[width=.9\linewidth]{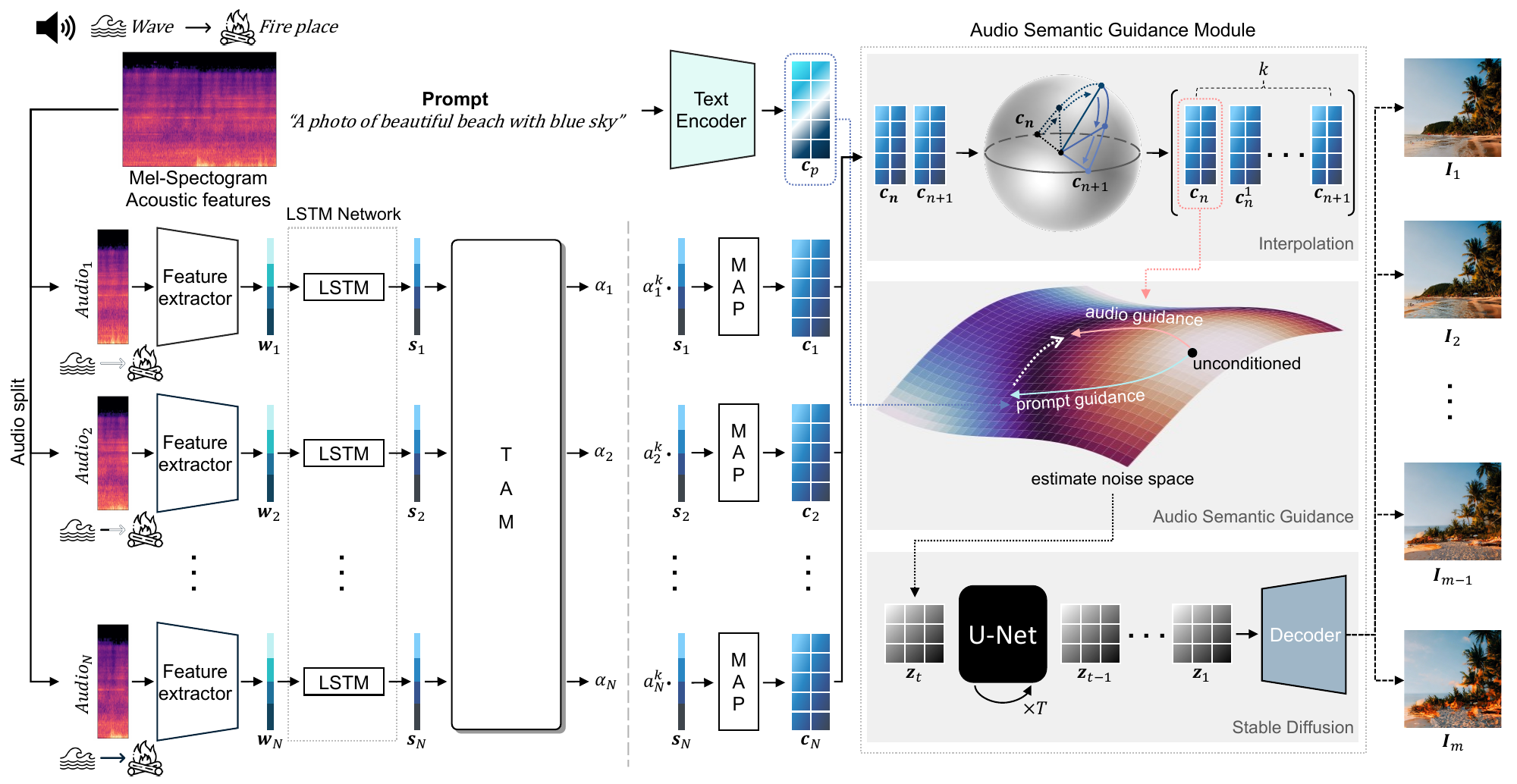} %
  \caption{An overview of our proposed TPoS model. Our model consists of two main modules: (i) Audio Encoder, which produces a sequence of latent vectors, encoding temporal semantics of audio input by utilizing CLIP~\cite{radford2021learning} space and highlighting the important temporal features and (ii) Audio Semantic Guidance Module, which is based on the diffusion process, generating video frames that are temporally consistent and audio-reactive.}
  \label{fig:overview}
  \vspace{-2em}
\end{center}
\end{figure*}


%% file: src/method.tex
\section{Method}

In this work, we propose a novel audio-driven video generation method, which generates video frames conditioned on audio sequences, ensuring temporal consistency between consecutive frames and temporal correspondence with audio inputs. As shown in Figure~\ref{fig:overview}, our model consists of two main parts: (i) {\em Audio Encoder}, which encodes temporal semantics of audio sequences, producing a sequence of the latent vectors (Section~\ref{sec:audio}). (ii) {\em Audio Semantic Guidance Module}, which uses the above-mentioned latent vectors as a condition in the diffusion process to generate corresponding image outputs, which are temporally consistent (by our identity regularizer) and audio-reactive (see Section~\ref{sec:video}).


\subsection{Preliminary: Latent Stable Diffusion}
Latent Diffusion Models (LDMs)~\cite{rombach2022high} are the method that uses an encoder to convert a noised latent vector $\mathbf{z}^T$ to a denoised latent vector $\mathbf{z} = \mathbf{x} +\mathbf{\epsilon}$, where $\mathbf{z}$ is a latent vector of an input image $\mathbf{x}$ and $\mathbf{\epsilon}$ is a noise. Stable Diffusion~\cite{rombach2022high} is part of a conditional generation model that can synthesize an image given a condition $\mathbf{y}$. It uses U-Net~\cite{ronneberger2015u} as denoising autoencoders represented by $\epsilon_\theta$.
To generate an output image, the autoencoder $\epsilon_\theta(\mathbf{z}^t, t, \tau_\theta(\mathbf{y}))$ takes three inputs: the noised latent vector $\mathbf{z}^t$, a sequence $t$ that is uniformly sampled from the set ${1, \ldots, T}$, and a conditional input $\mathbf{y}$.

Conditional input $\mathbf{y}$ is first transformed into a latent vector $\mathbf{c}_p$ through a pretrained function $\tau_\theta$ and then it is fed into a cross-attention layer of the U-Net as the key and value, which is then combined with $\mathbf{z}^t$ by an attention mechanism, where the query is the flattened intermediate representation of the U-Net. Next, the denoising autoencoder $\epsilon_\theta$ is used to denoise $\mathbf{z}^t$ from $t=T$ to $t=1$ sequentially. After $T$ denoising steps, the resulting denoised latent vector $\mathbf{z}~(=\mathbf{z}^1)$ is transmitted to the decoder, which produces the final output image $\tilde{\mathbf{x}}$.


\subsection{Encoding Temporal Semantics from Audio}~\label{sec:audio}
\begin{figure}[t]
    \begin{center}
        \includegraphics[width=\linewidth]{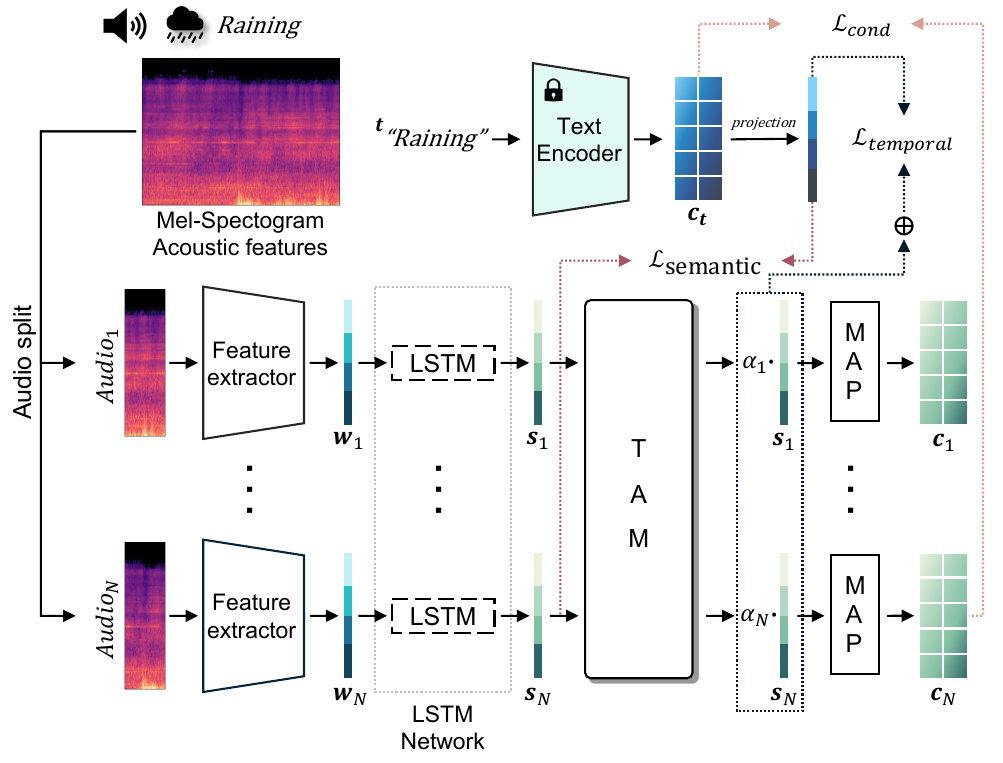} %
        \caption{An overview of our Audio Encoder training process. Our model generates temporally-encoded audio embeddings with an LSTM~\cite{hochreiter1997long} layer and Temporal Attention Module (TAM). Audio input is partitioned into $N$ segments, and each of these is encoded and used as a condition to manipulate audio-reactive video sequences (e.g. light rain $\rightarrow$ heavy rain). This is done by our Mapping Module (MAP), which maps the audio embedding to the latent space of Stable Diffusion.}
    \label{fig:audio}\vspace{-3em}
    \end{center}
\end{figure}

We use an audio modality as a source of generating temporal conditions. We divide the audio mel-spectogram into uniform snippets (segments) to extract abundant audio features and feed this into the Audio Encoder to extract both temporal semantic features and intensity of audio in end-to-end manner.  The figure of our training process is illustrated in Figure~\ref{fig:audio}.

\myparagraph{Audio Feature Extraction.}
Audio inputs are first transformed into a mel-spectrogram representation, denoted as $\mathbf{x}^{a} \in \mathbb{R}^{d \times w}$, where $d$ represents the number of mel-frequency bins and $w$ is the width of the spectrogram. To incorporate time information, the mel-spectrogram is divided into $N$ segments. Each segment, denoted as $\mathbf{x}_{n}^{a} \in \mathbb{R}^{d \times \lceil \frac{w}{N}\rceil }$, where $n\in\{1,\dots,N\}$, is then fed into a shared feature extraction module, i.e., the pre-trained ResNet18~\cite{he2016deep}. The feature extraction module $f_a(\cdot)$ learns to extract low-level features from each audio segment regardless of its time dependency, i.e., $\mathbf{w}_{n}=f_a(\mathbf{x}_{n}^{a})$ 




\myparagraph{LSTM-based Temporal Semantic Encoder.}
As our goal is to generate audio-reactive video frames, it is also important to encode temporal changes (or relations) of the given audio inputs. Similar to Lee~\etal~\cite{lee2022eccv}, given audio features $\mathbf{w}\in\{\mathbf{w}_1, \mathbf{w}_2, \dots, \mathbf{w}_N\}$, which encodes per-segment disjoint audio representation, we apply the standard Long Short-Term Memory (LSTM) network~\cite{hochreiter1997long} to encode temporal relations or changes between consecutive audio features $\mathbf{w}$. Formally, our LSTM takes the audio feature $\mathbf{w}_{n-1}$ as input and updates its hidden state $\mathbf{h}_{n}$, producing an output $\mathbf{s}_t$: i.e. $(\mathbf{s}_{n}, \mathbf{h}_{n}) = \texttt{LSTM}(\mathbf{h}_{n-1},\mathbf{w}_{n-1})$. These outputs $\mathbf{s}\in\{\mathbf{s}_1, \mathbf{s}_2,\dots, \mathbf{s}_N\}$ are then fed into Temporal Attention Module (TAM) to further encode temporal semantics. 




\myparagraph{Aligning Audio Semantics with Image-Text CLIP Joint Space.}
As we will use the output $\mathbf{s}$ as a condition to manipulate video frames, it is important to ensure those audio features are well-aligned with other text and visual features in the CLIP~\cite{radford2021learning}-based joint embedding space. Similar to Lee~\etal~\cite{lee2022cvpr}, given the pre-trained image-text CLIP space, we apply the following loss $\mathcal{L}^{a\leftrightarrow t}_{\text{CLIP}}$ with the InfoNCE loss~\cite{oord2018representation} $l_{sim}$ such that positive pairs (e.g. an audio of raining and a text prompt ``raining'') are pulled close to each other, while negative pairs are pushed farther away. 
\begin{equation} 
    \mathcal{L}^{a\leftrightarrow t}_{\text{CLIP}} =l_{sim}(\mathbf{s}_{N},\texttt{CLIP}_{t}(\mathbf{t}))+l_{sim}(\texttt{CLIP}_{t}(\mathbf{t}),\mathbf{s}_{N})
\end{equation}
where $\texttt{CLIP}_{t}$ is a pre-trained CLIP-based text encoder, which takes a text prompt $\mathbf{t}$ obtained from audio class labels as input, yielding an $d$-dimensional feature. Note that we only apply this loss for the final output $\mathbf{s}_{N}$ for efficient training. Given a set of positive pairs, we apply the following InfoNCE loss $l_{sim}(\mathbf{a}, \mathbf{b})$:
\begin{equation}
    l_{sim}= - \log \frac{\exp(\langle \mathbf{a}_i, \mathbf{b}_i\rangle )/\tau}{\sum_{j} \exp(\langle \mathbf{a}_i, \mathbf{b}_j\rangle )/\tau}
\end{equation}
where $\langle \mathbf{a}_i, \mathbf{b}_i\rangle$ represents the cosine similarity with temperature $\tau$. Note that we set $\tau$ to 0.07.

\myparagraph{Augmenting Audio Semantics.}
Audio data is often limited in volume and diversity; thus, augmentation techniques may be required to extract better-quality audio semantic features, preventing a representation collapse. We use SpecAugment~\cite{park2019specaugment} to apply random transformations (such as masking our certain frequency bands or time segments), yielding augmented audio inputs. We further add the InfoNCE loss~\cite{oord2018representation} $\mathcal{L}^{a\leftrightarrow a'}_{\text{CLIP}}$ to pull augmented audio features together.
\begin{equation}
    \mathcal{L}^{a\leftrightarrow a'}_{\text{CLIP}}= l_{sim}(\mathbf{s}_{N}, \mathbf{s}'_{N}) + l_{sim}(\mathbf{s}'_{N}, \mathbf{s}_{N}) 
\end{equation}
where apostrophe indicates augmented view of an original audio data. Finally, we use the semantic loss $\mathcal{L}_{\text{semantic}}$: i.e.  $\mathcal{L}_{\text{semantic}} = \mathcal{L}^{a\leftrightarrow t}_{\text{CLIP}} + \lambda_s\mathcal{L}^{a\leftrightarrow a'}_{\text{CLIP}}$ where $\lambda_{s}$ is set to 0.6.

\begin{figure}[t]
    \begin{center}
        \includegraphics[width=\linewidth]{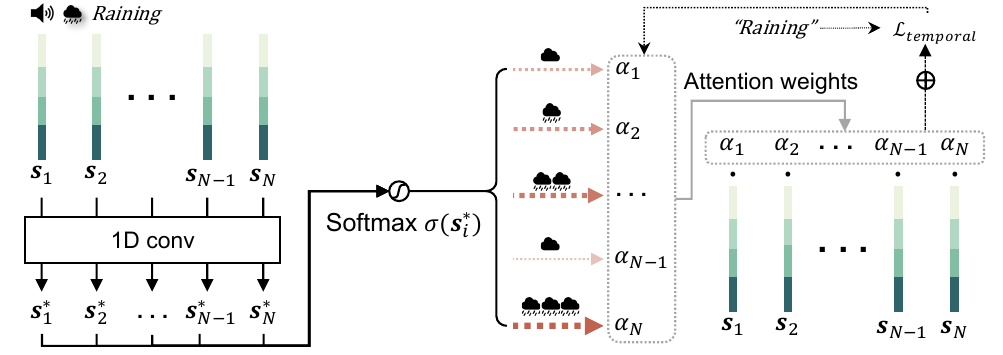} %
        \caption{Details of our Temporal Attention Module (TAM). Temporal Attention Module effectively captures the important audio segments (e.g. harsh raining sound) as attention weights, which are guided by CLIP~\cite{radford2021learning}'s text embedding (e.g. ``raining'') in training phase and express the magnitude of audio in test phase.}
    \label{fig:attention}\vspace{-2em}
    \end{center}
\end{figure}

\myparagraph{Temporal Attention Module (TAM).}
As shown in Figure~\ref{fig:attention}, we further use an attention-based module to encode temporal semantics from the audio inputs. We empirically observe that adding this module helps improve the quality of video frame generation, which is probably due to the fact that the model becomes flexible to focus on more important temporal information, improving the model's representation power. Formally, we first compute attention weight $\mathbf{\alpha}_{n}$ for a given audio feature $\mathbf{\mathbf{s}}_n$ by applying an MLP layer $f_{\text{proj}}$ followed by a softmax operation: i.e. $\mathbf{\alpha}_{n} = \exp(f_{\text{proj}}(\mathbf{s}_{n}))/\sum_{n} \exp(f_{\text{proj}}(\mathbf{s}_{n}))$ such that $\sum_n\mathbf{\alpha}_{n} = 1$. We compute the weighted sum of audio features based on attention weights, yielding an attended audio feature $\mathbf{o}^a=\sum_{n}\mathbf{\alpha}_{n}\mathbf{s}_{n}$. We add another InfoNCE loss $\mathcal{L}_\text{temporal}$ to align the audio features with text:
\begin{equation}
    \mathcal{L}_\text{temporal} = l_{sim}(\mathbf{o}^{a},\texttt{CLIP}_{t}(\mathbf{t}))+ l_{sim}(\texttt{CLIP}_{t}(\mathbf{t})),\mathbf{o}^{a})
\end{equation}
Lastly, we also minimize the MSE loss between text embeddings (before the projection layer) and the projected audio feature $\texttt{MAP}(\mathbf{o}^a)$: $\mathcal{L}_\text{cond}=||\mathbf{c}_t - \texttt{MAP}(\mathbf{o}^a)||_2^2$. Note that we exclude the feature for the \texttt{<SOS>} token. More details are provided in the supplemental material.









\begin{table*}[t]
    \centering
    \caption{Comparison of the quality of generated video frames with state-of-the-art audio-to-video generations in terms of IS~\cite{salimans2016improved}, FVD~\cite{unterthiner2018towards}, and CLIP~\cite{radford2021learning}-based distances.}
    \label{tab:Comparison2}
    \setlength{\tabcolsep}{10pt}
    \renewcommand{\arraystretch}{1.3} 
    \resizebox{0.85\linewidth}{!}{
        \begin{tabular}{@{}lccccc}\toprule
            Model  & Input & IS$\uparrow$ & FVD$\downarrow$ & CLIP$\uparrow$ (a $\leftrightarrow$ v) & CLIP$\uparrow$ (t $\leftrightarrow$ v) \\\midrule
            Sound2Sight~\cite{chatterjee2020sound2sight}  & 1st Frame & 1.02 $\pm$ 0.02 & 494.28 & 0.0364 &0.2164 \\
            CCVS~\cite{le2021ccvs}  & 1st Frame  & 1.30 $\pm$ 0.20 & 679.94 &0.1251 & 0.2360 \\\midrule
            Tr\"aumerAI~\cite{jeong2021traumerai}  & - & 1.47 $\pm$ 0.19 & 736.32 &0.1589&0.1778 \\
            Sound-guided Video Generation~\cite{lee2022eccv} & - & 1.16 $\pm$ 0.16 & 544.09 & 0. 1151 & 0.1702 \\\midrule
            Ours (w/o TAM)  & Latent Vector & 1.27 $\pm$ 0.23 & 483.76 & 0.1342 & 0.2370 \\
            Ours & Latent Vector & \textbf{1.49 $\pm$ 0.38} & \textbf{421.23} &\textbf{0.1964} & \textbf{0.2436} \\\bottomrule
        \end{tabular}\vspace{-1.0em}
        }
\end{table*}

\begin{figure*}[t]
\begin{center}
      \includegraphics[width=\linewidth]{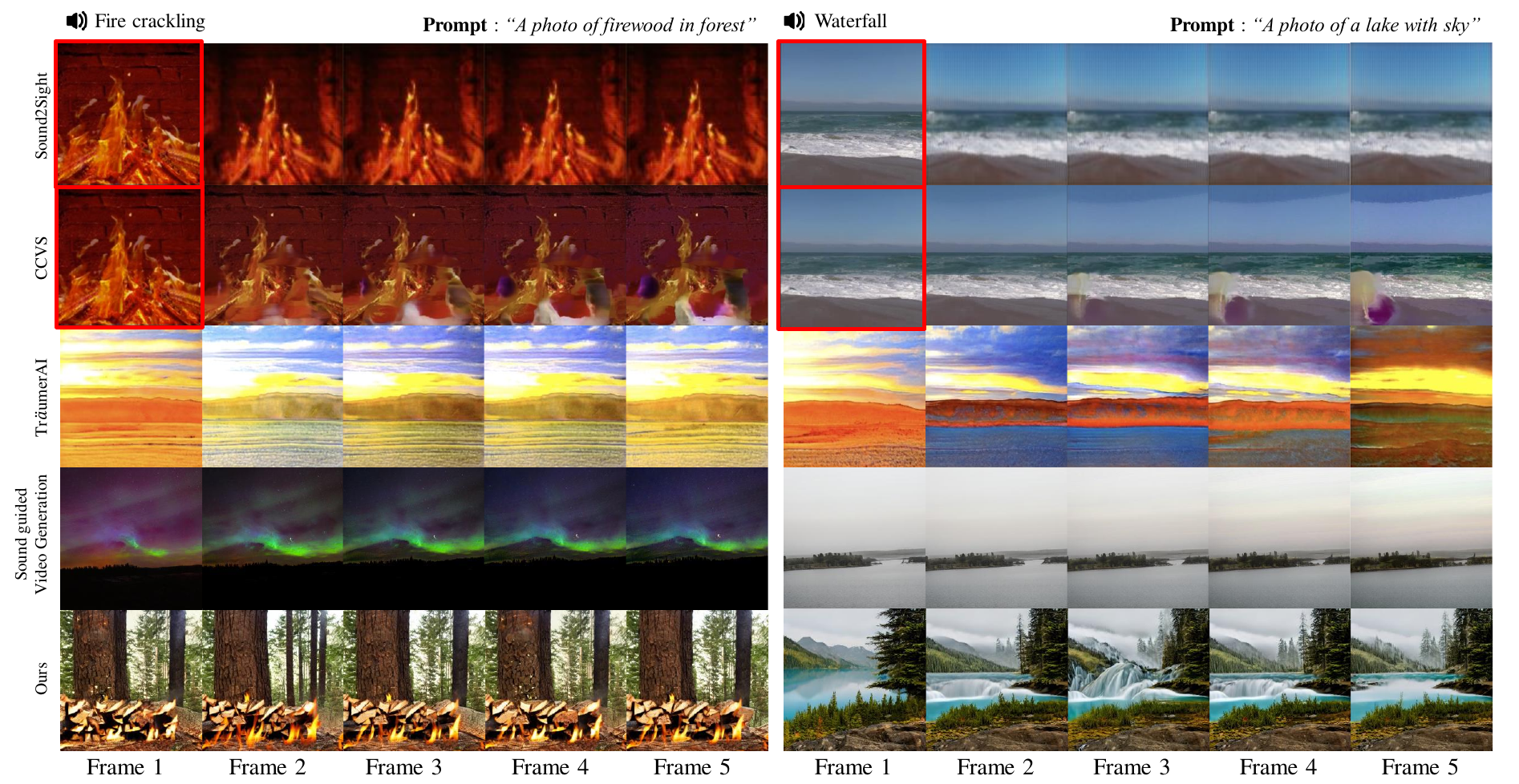}
  \caption{Examples of generated video frames (given fire crackling and waterfall audio) by Sound2Sight~\cite{chatterjee2020sound2sight}, CCVS~\cite{le2021ccvs}, Tr\"aumerAI~\cite{jeong2021traumerai}, Lee~\etal~\cite{lee2022eccv}, and ours. Note that Sound2Sight and CCVS use an initial frame (highlighted in a red box).}
  \label{fig:qualitative}\vspace{-1.5em}
\end{center}
\end{figure*}

\myparagraph{Total Loss.}
We train our Audio Encoder end-to-end by minimizing the following loss function $\mathcal{L}$:
\begin{equation}
\mathcal{L} = \mathcal{L}_\text{semantic} + \mathcal{L}_\text{temporal} + \mathcal{L}_\text{cond}
\end{equation}
%






\subsection{Generating Video Frames with Stable Diffusion} \label{sec:video}
\myparagraph{Initial Frame Generation from Text Prompt.}
Our model relies on leveraging combinational operations of denoising process. The estimated noise space can be either random or manual by visual input with the help of the diffusion encoder. Based on this guided diffusion, our model first generates the initial frame with a text prompt (e.g. ``a photo of beautiful beach with blue sky''). Given this generated image as \textit{content}, we manipulate its \textit{styles} according to audio inputs and generate corresponding video frames, i.e. given a series of latent vectors $\{\mathbf{c}_1, \mathbf{c}_2, \dots, \mathbf{c}_m\}$, we generate $m$ video frames. Note that the number of video frames is controllable by latent vector interpolation, as we will explain later. We follow the standard image generation process with the Stable Diffusion model~\cite{rombach2022high}, i.e., we compute a latent vector $\mathbf{c}_{p}$ in a CLIP-based embedding space given a text prompt, conditioning it to generate an image.



%

\myparagraph{Audio Semantic Guidance.}
We employed the SEGA~\cite{brack2022stable} framework to create video frames that incorporate sound style while preserving the main content identity. We first utilize the combination of attention $\mathbf{\alpha}_n$ and audio feature $\mathbf{s}_n$ with normalization of the scale of output feature by multiplying $N$ to produce output $\mathbf{c}_n$: i.e., $\mathbf{c}_n=N^k\mathbf{\alpha}_n^{k}\mathbf{s}_n$, where $k$ is hyper parameter that regulates attention ratio and is set to 1. The Audio Semantic Guidance module generates the $n$-th video frame by taking audio condition $\mathbf{c}_n$ as input to guide the diffusion models. The denoising autoencoder $\epsilon_\theta$ is executed $\delta-1$ times out of $T$ time to form incomplete noise along with the original text prompt meaning. From $t=\delta$, the audio semantic guidance operates through the following equation:
\vspace{-.5em}
\begin{equation}
    \begin{aligned}
        \tilde{\epsilon_{\theta}}(\mathbf{z}^{\delta} ,\mathbf{c}_p) := &\epsilon_{\theta}(\mathbf{z}^\delta,\mathbf{c}_{\varnothing}) + g(\epsilon_{\theta}(\mathbf{z}^{\delta},\mathbf{c}_p)- \epsilon_{\theta}(\mathbf{z}^{\delta},\mathbf{c}_{\varnothing})) \\ 
        &+ \lambda(\mathbf{z}^{\delta},\mathbf{c}_n)
    \end{aligned}
\end{equation}
where $\mathbf{z}^\delta$ is denoised latent vector at $t=\delta$, $g$ is the guidance scale of the text prompt, $\lambda(\mathbf{z}^\delta,\mathbf{c}_n)$ is the audio semantic guidance term, and $\mathbf{c}_\varnothing$ represents an unconditioned prompt that does not make any semantic difference. As a result, only the $\lambda(\mathbf{z}^\delta,\mathbf{c}_n)$ term has been added to the original denoising process from $t=\delta$ to $t=1$. Note that $\mathbf{z}^T$ is fixed through frames in one video.  
The audio semantic guidance $\lambda(\mathbf{z}^\delta,\mathbf{c}_n)$ is defined as follows:

\begin{equation}
    \lambda(\mathbf{z}^\delta,\mathbf{c}_n)=  g_s(\epsilon_{\theta}(\mathbf{z}^{\delta},\mathbf{c}_n) - \epsilon_{\theta}(\mathbf{z}^{\delta}_{\varnothing},\mathbf{c}_\varnothing)) + 
    \sigma_{m}\Phi_{m}
\end{equation}
%
where $\sigma_{m} \in [0,1]$ is  the momentum hyper parameter that scales the momentum $\Phi_{m}$ and $g_s$ focuses on the relevant dimensions to the audio manipulation task (More details can be found in the supplemental material or SEGA~\cite{brack2022stable}). Different from SEGA framework, we only consider positive guidance of audio semantics, which is the semantic difference between the guidance provided by $\mathbf{c}_n$ and the unconditioned guidance $\mathbf{c}_\varnothing$. The generation process conditioned by $\mathbf{c}_n$ is separately working through diffusion processes so that the different semantic meaning or magnitude which presents in $\mathbf{c}_{n}$ can generate frames independently.

\vspace{-1em}
\myparagraph{Temporal Frame Interpolation.}
We use an interpolated latent vector to generate continuous video frames between two consecutive frames. Following the work by Ramesh \etal.~\cite{ramesh2022hierarchical}, we apply a spherical linear interpolation between all consecutive pairs of $\mathbf{c}_n$ and $\mathbf{c}_{n+1}$, yielding $k$ interpolated latent vectors. These vectors are then used as a condition for the diffusion models to generate temporally-interpolated video frames.

%% file: src/experiments.tex
\section{Experiments}
\myparagraph{Datasets.}
We use two Audio-Video datasets to train our Audio Encoder: VGG-Sound~\cite{chen2020vggsound} and Landscape~\cite{lee2022eccv}. VGG-Sound is an audio dataset with about 170,000 of 10-second clips of audio-video data, which consists of 309 classes. The dataset has numerous `in the wild' audio data that spans a large number of challenging acoustic environments and real application noise characteristics. Since audio of nature sound is the perfect tool to stylize compared to the class such as people talking or sports, we add about 9,000 audio clips of Landscape audio dataset in the training process. For obtaining test sets, class-balanced sampling is applied to Landscape dataset. 

\myparagraph{Baselines.}
We compare our methods with existing audio-to-video generation methods, Sound2Sight~\cite{chatterjee2020sound2sight}, CCVS~\cite{le2021ccvs}, and StyleGAN~\cite{skorokhodov2022stylegan} based Tr\"aumerAI~\cite{jeong2021traumerai} and Sound-guided Video Generation~\cite{lee2022eccv}. We follow experiment methods by Lee \etal~\cite{lee2022eccv}. All baselines are trained or finetuned with Landscape dataset~\cite{lee2022eccv}. Since Sound2Sight and CCVS need first frame to generate the video, we randomly select first frame from the Landscape dataset at the inference task. For Tr\"aumerAI and Sound-guided Video Generation, we first pre-train StyleGAN~\cite{karras2021alias} on high fidelity benchmark datasets (LHQ datasets~\cite{skorokhodov2021aligning}) and then train the model to navigate the latent space by Landscape dataset. Note that randomly initialized vector for Tr\"aumerAI~\cite{jeong2021traumerai} and Sound-guided Video Generation~\cite{lee2022eccv} are given. 
For our model, we initially generate image from random noise space and randomly sample the prompt related with landscape to generate the landscape like video for fair comparison.


\begin{figure}[t]
\begin{center}
      \includegraphics[width=\linewidth]{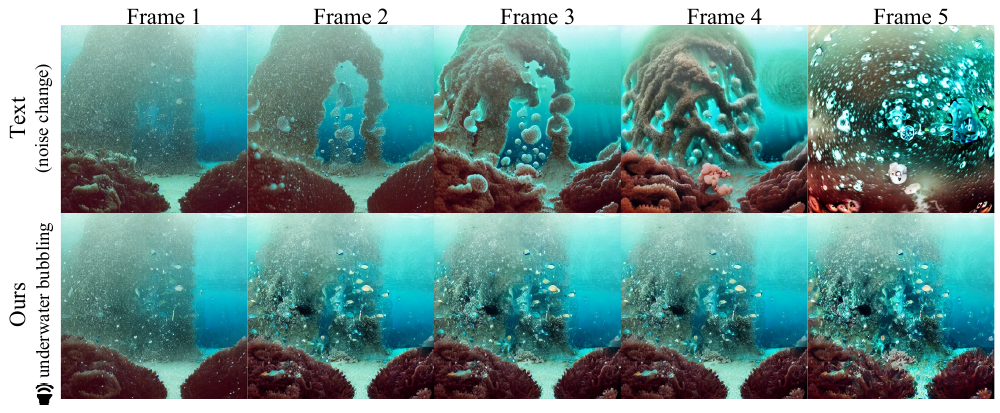} %
  \caption{Examples of our video generation conditioned on text prompt (top) and audio (bottom).}
  \label{fig:vsText}
  \vspace{-7mm}
\end{center}
\end{figure}

\myparagraph{Evaluation Metrics.}
We use the following two video quality metrics for our evaluations: (i) Fr\'echet Video Distance (FVD)~\cite{unterthiner2018towards} and (ii) Inception Score (IS)~\cite{salimans2016improved}. FVD is used to assess video quality by measuring the distribution gap between real and synthesized videos in the latent space. Additionally, IS score is commonly employed to evaluate the effectiveness of Generative Adversarial Networks (GANs) by computing the KL-divergence between the label distribution of each image and the marginal label distribution. To implement these, we fine-tune Inflated 3D ConvNet~\cite{carreira2017quo} with Landscape~\cite{lee2022eccv} dataset for FVD and used pre-trained InceptionNet~\cite{kay2017kinetics} that is trained on ImageNet~\cite{deng2009imagenet} dataset. We also measure CLIP~\cite{radford2021learning}-based cosine similarity (CLIP score) between audio and image as well as text and image. To obtain the textual pivot feature, we fed the following prompt ``The photo of \texttt{<class>}'' into CLIP text encoder. 


\myparagraph{Quantitative Experiments.}
For fair comparison with other existing baselines such as Tr\"aumerAI~\cite{jeong2021traumerai} and Sound-guided Video Generation~\cite{lee2022eccv} which does not get a hint about what to generate, we use prompt that originally does not generate the style of sound. We set fps 20 and generated videos to extract images from all baselines. Table~\ref{tab:Comparison2} shows that our approach produces the best quality results as video. Additionally, to ensure that the generated videos are semantically related to the sound, we compare the cosine similarity between text-audio and video embedding. Our methods shows a superior performance in terms of multi-modal semantics.

\myparagraph{Qualitative Video Quality Comparison.}
We first evaluate the quality of generated video frames. In Figure~\ref{fig:qualitative}, we provide typical examples of generated video frames by (from top) Sound2Sight~\cite{chatterjee2020sound2sight}, CCVS~\cite{le2021ccvs}, Tr\"aumerAI~\cite{jeong2021traumerai}, Lee~\etal~\cite{lee2022eccv}, and ours. Note that Sound2Sight~\cite{chatterjee2020sound2sight} and CCVS~\cite{le2021ccvs} need the initial frame as input (see red boxes). With audio inputs, such as fire crackling and waterfall, as a condition, ours generally generate temporally-consistent audio-reactive video frames. We observe that Sound2Sight~\cite{chatterjee2020sound2sight} and CCVS~\cite{le2021ccvs} often show blurring artifacts, while StyleGAN-based Tr\"aumerAI~\cite{jeong2021traumerai} and Sound-guided Video Generation~\cite{lee2022eccv} often fail to generate audio-reactive video frames, e.g., they produce landscape scenes that are not semantically aligned with a fire crackling audio. However, ours generate a scene of a waterfall or a fire on firewood, aligning well with audio inputs.



\begin{figure}[t]
\begin{center}
      \includegraphics[width=\linewidth]{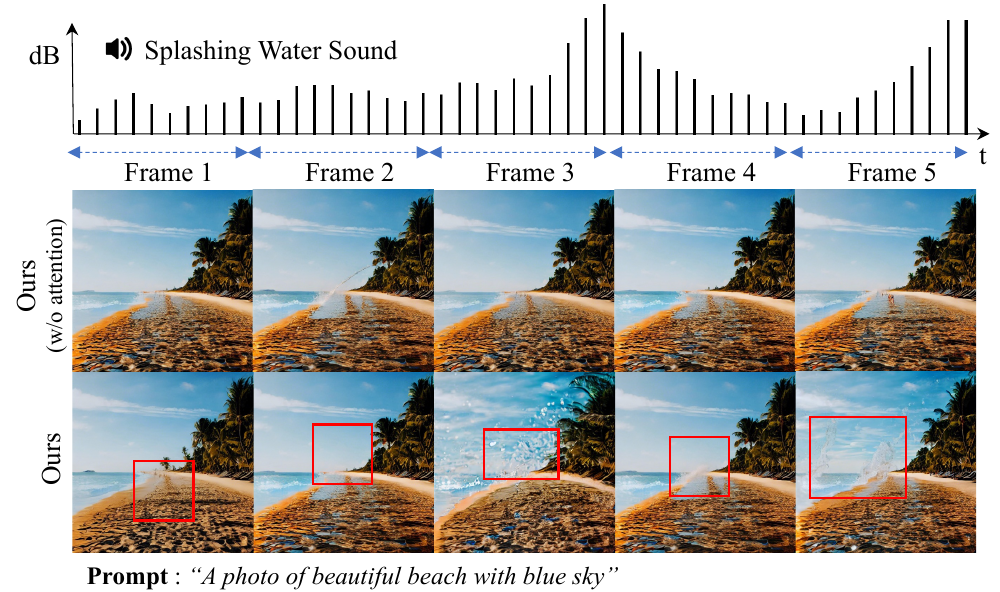} %
  \caption{Generated video frames \textit{with} and \textit{without} our Temporal Attention Module. We generate video frames with splashing water sound where its amplitude changes over time (see top).}\vspace{-2em}
  \label{fig:ablation}
\end{center}
\end{figure}

\myparagraph{Comparison to Video Generation with Text.}
To analyze the benefit of audio modality, we conduct a qualitative experiment to compare the effect of audio and text modalities in generating visual content. As shown in Figure~\ref{fig:vsText}, we first generate an initial frame with the text prompt ``A photo of deep in the sea.'' Then we generate the next frames with text ``underwater bubbling'' (top) and underwater bubbling sound (bottom). It is difficult to make temporal changes conditioned on text unless we train our model with a text-video dataset. Thus, we instead change the noise scale to make temporal changes, preserving identity. However, as shown in Figure~\ref{fig:vsText} (top), it generates distortions or a linear change, but this is not the case for audio. Our model with audio generates visually-appealing video frames.

\myparagraph{Effect of Temporal Attention Module.}
We use Temporal Attention Module (TAM) to improve the representation power to encode temporal semantics better. To analyze this, we perform an ablation study with and without TAM to see its effect on video frame generation. We observe in Figure~\ref{fig:ablation} that our model is indeed able to generate video frames reactive to the audio changes over time (compare the changes along with the given splashing water sound).



\begin{figure}[t]
\begin{center}
\includegraphics[width=\linewidth]{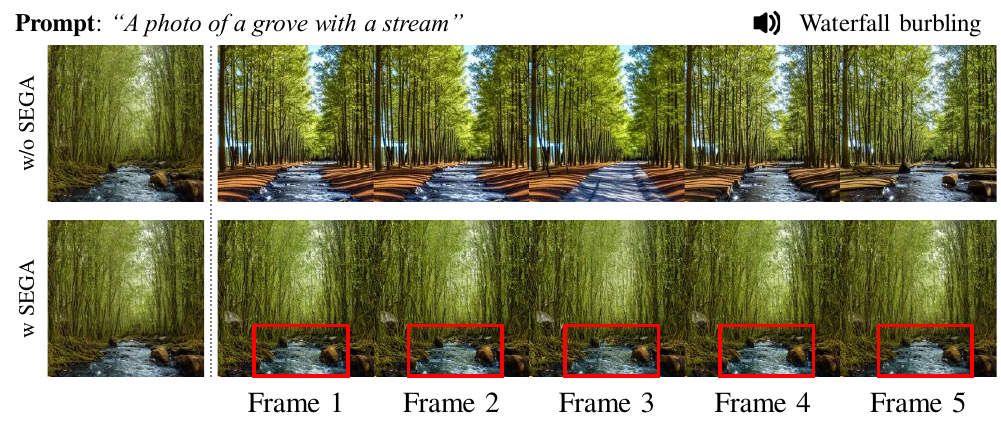} %
  \caption{Generated video frames with and without our Audio Semantic Guidance. Edited parts by Audio Semantic Guidance are highlighted with red box (see second row).}\vspace{-2em}
  \label{fig:sup_sega}
\end{center}
\end{figure}

\begin{figure}[t]
\begin{center}
\includegraphics[width=\linewidth]{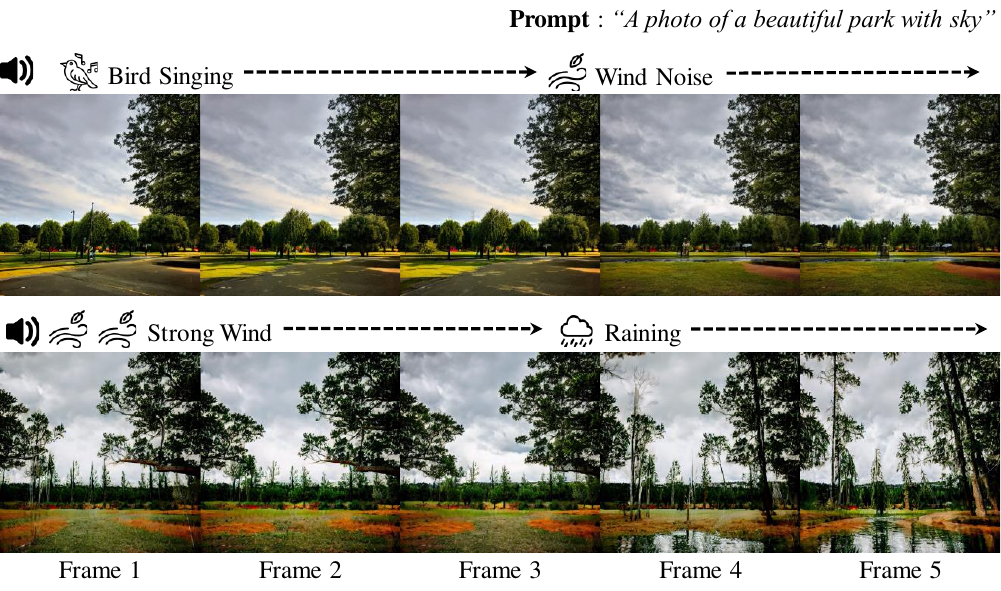} %
  \caption{Examples of generated video frames with a sound that changes over time (e.g. bird singing $\rightarrow$ wind noise).}
  \label{fig:transition}
  \vspace{-2em}
\end{center}
\end{figure}

\myparagraph{Effect of Audio Semantic Guidance.}~\label{sec:guidance}
We demonstrate the effect of Audio Semantic Guidance with ablation study. First, we generate video frames without Audio Semantic Guidance by replacing audio semantic guidance term $\lambda(\mathbf{z}^{\delta},\mathbf{c}_n)$ with $\epsilon_{\theta}(\mathbf{z}^{\delta},\mathbf{c}_n)$ (refer the notation in Section~\ref{sec:video}). To remove the effect of Audio Semantic Guidance, we also set $\delta$ to $T$. As illustrated in Figure~\ref{fig:sup_sega} (top),
without Audio Semantic Guidance, the model has a tendency to produce unnecessary changes (e.g. grass $\rightarrow$ toil) and struggles to achieve consistent alignment with audio (e.g. producing a road with a waterfall burbling sound). On the other hand, by leveraging Audio Semantic Guidance as in Figure~\ref{fig:sup_sega} (bottom), we can generate sequential video frames that not only has consistent \textit{content} but also represents natural temporal variations according to audio sound (e.g. producing wave of water with waterfall burbling sound), resulting in enhanced naturalness.




\myparagraph{Experiments of Semantic Transition.}
In Figure~\ref{fig:transition}, we provide examples where audio inputs are changed (e.g., bird singing $\rightarrow$ wind noise, strong wind $\rightarrow$ raining). As we observe in that figure, our model successfully adapts to the audio change, generating video frames accordingly. This may confirm that our model is indeed conditioned on the audio sequence and can generate audio-adaptive video frames.

\begin{figure}
\begin{center}
      \includegraphics[width=\linewidth]{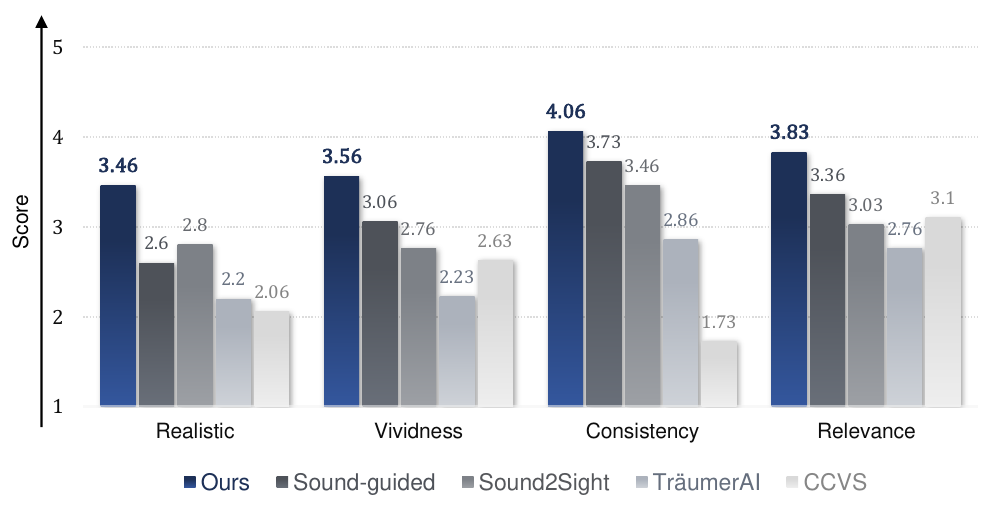} %
        \caption{Our human evaluation results. We conduct a user study with 100 participants on Amazon Mechanical Turk~(AMT). Participants are shown generated video frames and asked to evaluate them in terms of realistic, vividness, consistency, and relevance. The Likert scale is used (higher is better).}
  \label{fig:userstudy}
  \vspace{-5mm}
\end{center}
\end{figure}

\myparagraph{User study.}
Further, we conduct a human evaluation to evaluate the video quality by human judges. We recruit 100 participants from Amazon MTurk. Participants are shown video frames generated by five different audio-driven video generation models: Sound2Sight~\cite{chatterjee2020sound2sight}, CCVS~\cite{le2021ccvs}, TräumerAI~\cite{jeong2021traumerai}, Lee~\etal~\cite{lee2022eccv}, and ours. Participants are asked to evaluate the given video frames in terms of realism, vividness, consistency, and relevance. We use a five-point Likert scale where ideal video frames will get all five points. More details are provided as supplemental material. We observe in Figure~\ref{fig:userstudy} that our proposed method outperforms the other approaches in all categories. These results are consistent with our quantitative and qualitative results.

\begin{figure}[t]
\begin{center}
\includegraphics[width=.9\linewidth]{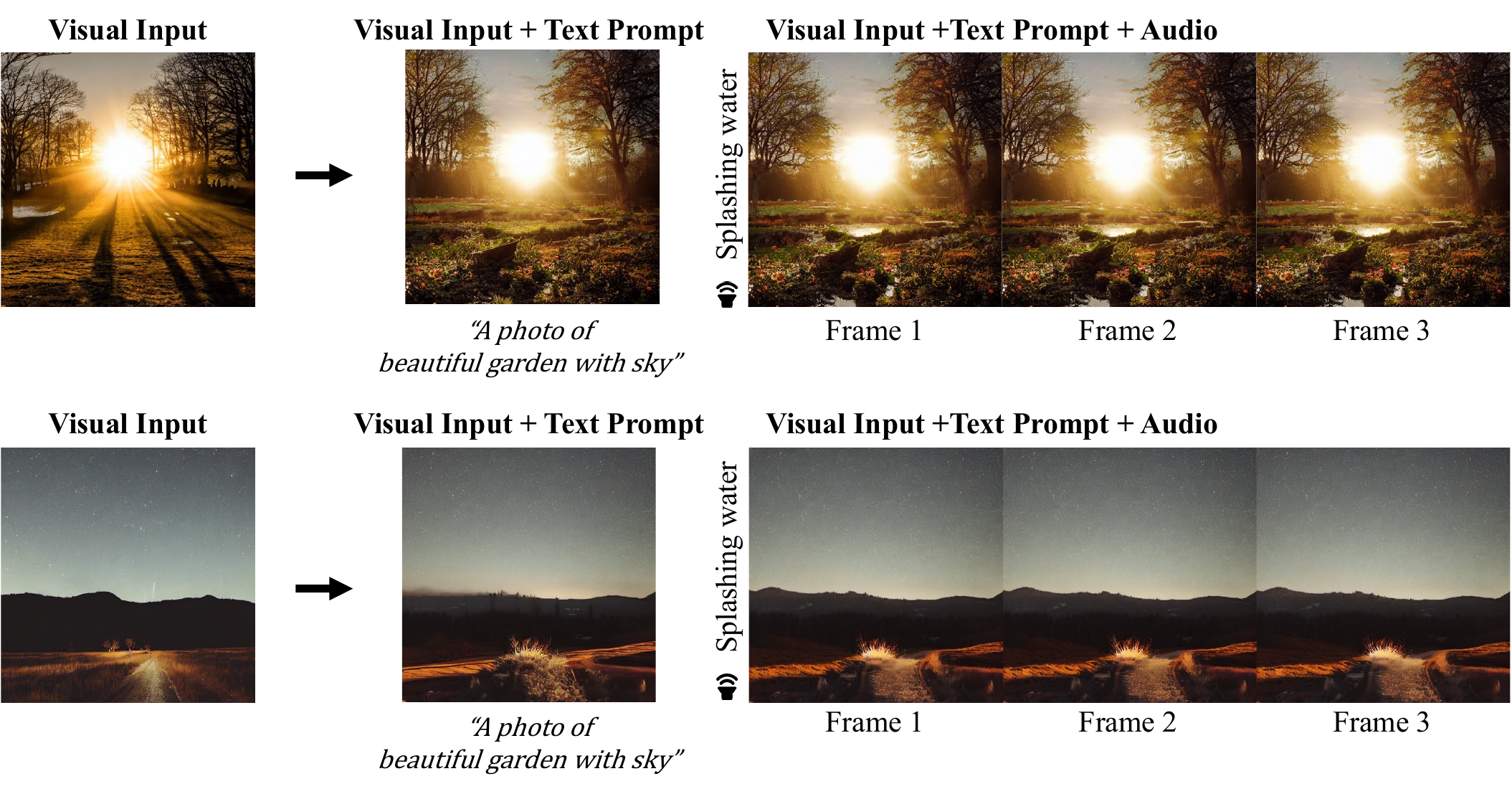} %
 \vspace{-3mm}
 \caption{Example of generated video with visual input, text prompt and audio sound. (e.g., 1st row: sun light photo conditioned by text ``A photo of beautiful garden with sky'' and splashing water sound.)}
  \label{fig:visual_example}
  \vspace{-4mm}
\end{center}
\end{figure}

\myparagraph{Visual Conditioning.}
Our model is capable of producing an initial frame using visual input (such as images) with the trained diffusion encoder. Figure~\ref{fig:visual_example} demonstrates that our framework leverages latent space of a diffusion model with a combination of visual input, text prompt and audio. We first randomly choose visual input (e.g., sun light photo) and generate an initial frame with the text prompt (e.g., ``A photo of beautiful garden with sky''). Then we incorporate an audio sound (e.g, splashing water) to generate videos. Shown in Figure~\ref{fig:visual_example}, it can produce videos featuring a garden with water where the sun is centered (1st row) or where a black mountain in the background (2nd row).


\begin{figure}[t]
\begin{center}
      \includegraphics[width=\linewidth]{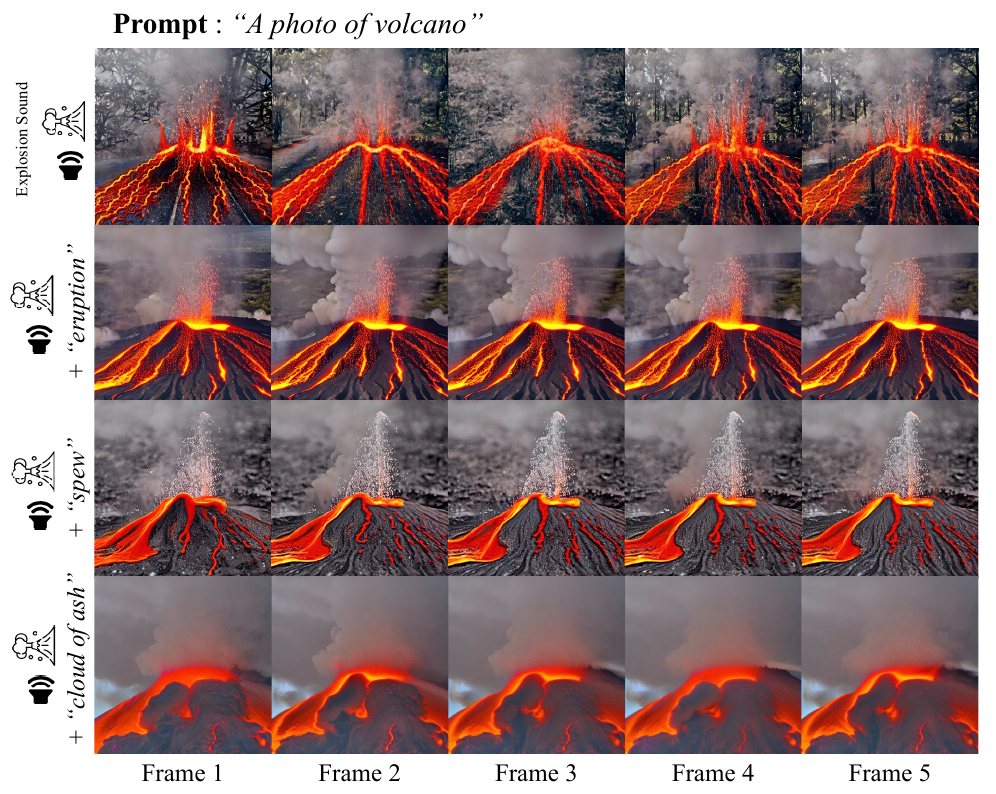} %
  \caption{Example of generated videos with audio-text joint condition (e.g., 2nd row: conditioned with text ``eruption'' and explosion sound)}
  \label{fig:application}
  \vspace{-2em}
\end{center}
\end{figure}

\myparagraph{Text-Audio Joint Conditioning.}
As our model is built upon the Stable Diffusion model, it is also possible to use text and audio as a condition together. In Figure~\ref{fig:application}, we provide an example where we generate video frames conditioned on a sound of an explosion along with texts, such as ``eruption'', ``spew'', or ``cloud of ash.'' (see 2nd-4th rows) Preserving temporal semantics, our model successfully generates video frames guided by text as well.  

\begin{figure}[t]
\begin{center}
\includegraphics[width=\linewidth]{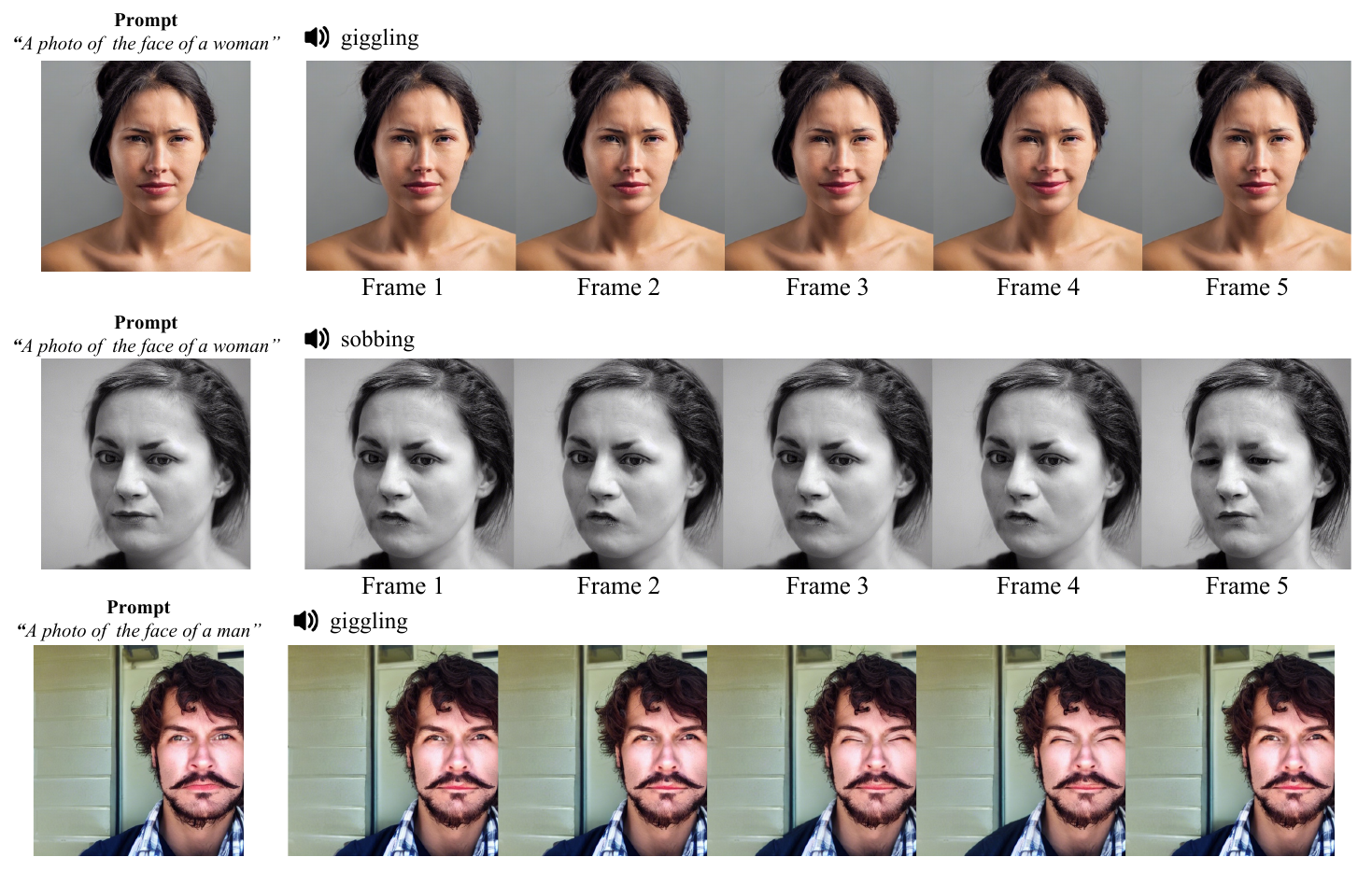} \vspace{-2mm}
  \caption{Examples of face generation with our methods. We use giggling and sobbing sound to manipulate face expressions of human face.}
  \label{fig:face}
  \vspace{-3em}
\end{center}
\end{figure}

\myparagraph{Face Generation.}
 Our Audio Semantic Guidance enables detailed adjustment in certain areas, such as face generation. Figure~\ref{fig:face} shows examples of human faces conditioned on giggling sound (1st and 3rd rows) and sobbing sound (2nd row). Our model excels in preserving the core facial attributes while manipulating emotional expressions.

%% file: src/conclusion.tex
\section{Conclusion}

In this paper, we propose The Power of Sound (TPoS), a novel audio-to-video generation with Stable Diffusion. Our work extends the usage of audio modality on generation models, and broaden the methods of using Stable Diffusion by generating realistic videos by our Audio Encoder. Superior performances are achieved over widely-used audio-to-video benchmarks, reflected by objective evaluations and User study, hence attributing towards the new formulation of video generation with audio modality.